\documentclass[12pt]{iopart}
\usepackage{iopams} 
\usepackage[dvips]{graphicx,color}
\usepackage[latin1]{inputenc}

\begin{document}

\title[Activity, diffusion, and correlations in a 2-D conserved stochastic
sandpile]{Activity, diffusion, and correlations in a two-dimensional conserved
stochastic sandpile}

\author{Sharon Dantas da Cunha$^{1}$, Luciano Rodrigues da Silva$^{2,3}$,
Gandhimohan M. Viswanathan  $^{2}$ and Ronald Dickman$^{4,5}$}
\address{$^{1}$Escola de Ci\^encias e Tecnologia, Universidade Federal do Rio
Grande do Norte, Campus Universit\'ario, 59078-970 Natal, Rio Grande do Norte,
Brazil}
\address{$^{2}$Departamento de F\'isica Te\'orica e Experimental,Universidade
Federal do Rio Grande do Norte, Campus Universit\'ario, 59078-900 Natal, Rio
Grande do Norte, Brazil}
\address{$^3$National Institute of Science and Technology of Complex Systems,
Universidade Federal do Rio Grande do Norte, Campus Universit�rio, 59078-900
Natal, Rio Grande do Norte, Brazil}
\address{$^4$Departamento de F\'isica, ICEx, Universidade Federal de Minas
Gerais, Caixa Postal 702, 30161-970 Belo Horizonte, Minas Gerais, Brazil}
\address{$^5$National Institute of Science and Technology of Complex Systems,
Caixa Postal 702, 30161-970 Belo Horizonte, Minas Gerais, Brazil}
\eads{sharondantas@ect.ufrn.br, luciano@dfte.ufrn.br, gandhi@dfte.ufrn.br and
dickman@fisica.ufmg.br}

\date{\today}

\begin{abstract}

We perform large-scale simulations of a two-dimensional restricted-height
conserved stochastic sandpile, focusing on particle diffusion and
mobility, and spatial correlations. Quasistationary (QS) simulations yield the
critical particle density to high precision [$p_c =  0.7112687(2)$], and show
that the diffusion constant scales in the same manner as the activity density,
as found previously in the one-dimensional case. Short-time scaling is
characterized by subdiffusive behavior (mean-square displacement $\sim t^\gamma$
with $\gamma < 1$), which is easily understood as a consequence of the initial
decay of activity, $\rho(t) \sim t^{-\delta}$, with $\gamma = 1- \delta$.
We verify that at criticality, the activity correlation function $C(r)
\sim r^{-\beta/\nu_\perp}$, as expected at an absorbing-state phase transition.
Our results for critical exponents are consistent with, and somewhat more
precise than, predictions derived from the Langevin equation for stochastic
sandpiles in two dimensions.

\end{abstract}

\pacs{05.70.Ln, 05.50.+q, 05.65.+b}
\maketitle

\section{Introduction}

Sandpile models are the best known examples of self-organized criticality (SOC)
\cite{btw1,btw2,jensen,pruessner,dhar99}, in which the dynamics of a system
forces it to the critical point of an absorbing-state phase transition
\cite{bjp,granada} leading to scale-invariance in the apparent absence of
tunable parameters \cite{ggrin}. The SOC state can be attributed to the presence
of two well separated time scales \cite{ggrin,pre2000,vz1,vz2}, one
corresponding to the external energy input or driving force, and the other to
the microscopic evolution (e.g., avalanches). The separation between the two
times scales (also called {\it slow driving}) effectively tunes the system to
the neighborhood of an absorbing-state phase transition; the latter transition
is reached in the usual manner, by adjusting a control parameter, in models
known as {\it conserved sandpiles} (CS) or {\it fixed-energy sandpiles}
\cite{bjp,tb88,pmb96,dvz,vdmz}. The CS has the same local dynamics as the
corresponding driven sandpile, but a fixed number of particles. It is
characterized by a nonconserved order parameter (the activity density) which is
coupled to a conserved field \cite{rossi} whose evolution is arrested in
space-time regions devoid of activity.

In recent years, a number of studies have characterized the critical properties
of conserved stochastic sandpiles (CSS). As is usual in studies of critical
phenomena, theoretical discussions of scaling and universality are anchored in
the analysis of a continuum field theory or of a Langevin equation (i.e., a
nonlinear stochastic partial differential equation) that reproduces the phase
diagram and captures the fundamental symmetries and conservation laws of the
system. In the case of CSS, these symmetries and conservation laws define the
{\it conserved directed percolation} (CDP) universality class \cite{rossi}.
Extensive numerical studies of a Langevin equation corresponding to CDP are
described in \cite{ramasco,dornic}. The critical exponent values reported in
\cite{ramasco} are in good agreement with simulations of conserved lattice gas
models \cite{pastor,kockelkoren}, which exhibit the same symmetries and
conservation laws as CSS. There is now strong evidence that the CSS belongs to
the CDP universality class \cite{pre2006,bona2008}, although the existence of
this class has been questioned by Basu {\it et al.} \cite{basu}. According to
these authors, the CSS belongs to the usual directed percolation (DP) class;
further studies are required to verify this assertion.

Most studies of CSS characterize the critical region using the order parameter
(the activity density $\rho$, i.e., the fraction of active sites)
\cite{pre2000,pre2006,mnrst}. In this work we study the diffusion particles in a
two-dimensional CSS to characterize the phase transition through the diffusion
constant $D$, defined via the mean-square particle displacement. Dhar and
Pradhan suggested $D$ and the order parameter would be proportional in Abelian
sandpiles \cite{dhar2004,pradhan2006}. An earlier study by some of the present
authors verified that $D$ scales in the same manner as the activity density in
the stationary regime in one-dimensional stochastic sandpiles \cite{epjb2009}.
Dimensionality appears to play a nontrivial role in defining the universality of
CSS \cite{basu,manna1d}, in particular, different relations between CSS and DP,
and between CSS and an elastic interface model, in one and two dimensions.
Thus it is of interest to verify the relation between $D$ and $\rho$ in the
two-dimensional case as well.  In addition, we study (for the first time, to
our knowledge), the static activity correlation function and the particle
mobility in the CSS.

The remainder of this paper is organized as follows. In Sec. 2 we define the
models. In Sec. 3 we report simulation results for the diffusion constant and
the order parameter, and perform scaling analyses to extract estimates for the
critical exponents. We close in Sec. 4 with a summary of our results.

\section{Model}

We study modified conserved stochastic sandpiles, related to Manna's model
\cite{manna1,manna2}, called the {\it restricted-height sandpile} in the
standard version studied in \cite{pre2006,mnrst,epjb2009}. In addition we study
the effect of a weak force or ``drive" $f$ directed along one of the lattice
directions to determine the particle mobility. The model is defined on square
lattice of $L \times L$ sites, with periodic boundaries; the configuration is
specified by the number of particles, $z_{i,j} =$ 0, 1, or 2, at each site
$(i,j)$. Sites with $z_{i,j}=2$ are {\it active}, while those with $z_{i,j} \leq
1$, are said to be {\it inactive}. No site may harbor more than two particles.

The temporal evolution consists of a series of {\it toppling} events, in which
two particles are transferred from an active site to one or more of its first
neighbors $[(i - 1,j),(i + 1,j),(i,j - 1),(i,j + 1)]$ with equal probabilities.
The target sites for the two particles are chosen independently. If a particle
attempts to jump to a site already bearing two particles, it returns to the
toppling site. The evolution follows a continuous-time Markovian dynamics in
which each active site has a transition rate of unity to topple. At each step of
the evolution, one of the current $N_a$ active sites is chosen at random to
topple; the time increment associated each step is $\Delta t = 1/N_{a}$. In this
way, each active site waits, on average, one time unit before toppling.

In conserved sandpiles, the particle density, $p = N/L^d$, serves as a
temperature-like control parameter \cite{bjp}. Below the critical value, $p_c$,
the system eventually reaches an absorbing configuration ($N_a = 0$).  For $p >
p_c$, by contrast, the activity continues indefinitely ($N_a > 0$), in the
infinite-size limit. The order parameter associated with the phase transition is
the activity density, given by the fraction of active sites, $\rho = N_{a}
/L^2$. Although activity {\it must} continue indefinitely if $p > 1$,
remarkably, $p_c$ is in fact well below unity. For the model studied here, the
best estimate for the critical density is $p_c = 0.7112687(2)$, as shown below.

\section{Simulation method and results}

In this section we discuss our results, first for the usual restricted-height
sandpile, and then for the model under a weak drive. In both cases, we study
system sizes $L=$ $32$ to $2048$. Random numbers are generated using the GNU
Scientific Library for C++ \cite{gsl}. Initial configurations are generated by
inserting $N$ particles randomly on a square lattice of $L \times L$ sites,
subject to the restriction $z_{i,j} \leq 2$.

\subsection{Locating the critical point}

\begin{figure}[htb]
\begin{center}
\vspace{0.5cm}
\includegraphics[scale=0.29]{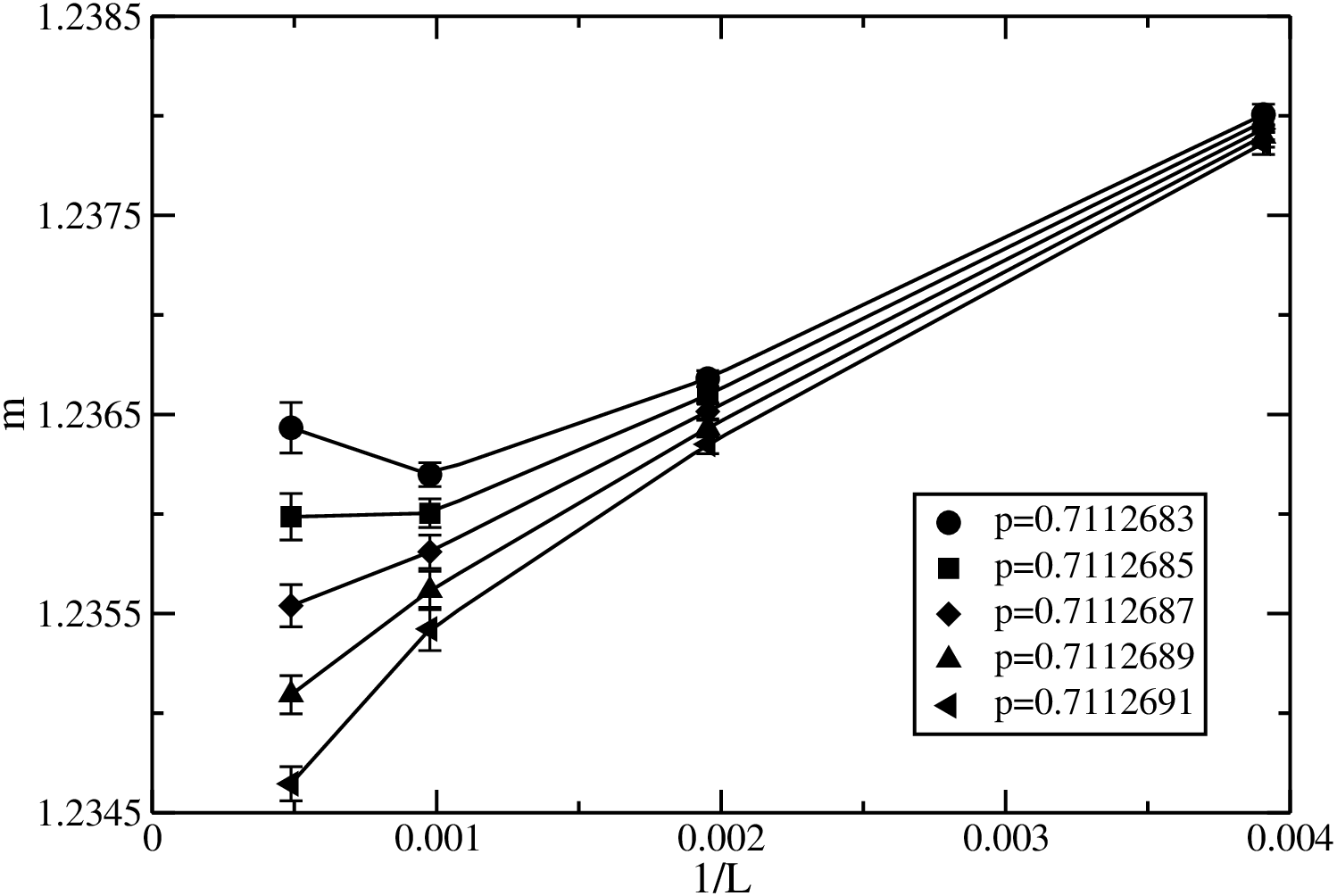}
\caption{\label{fig:m} Moment ratio m versus system size for particle
densities in the immediate vicinity of $p_c$.}
\end{center}
\end{figure}

We performed three sets of studies: two in the immediate vicinity of critical
point $p_c$, and a third in the supercritical region. Initially, we determine
the value of $p_c$ using the moment ratio $m = <\rho^2>/\rho^2$ in
quasistationary (QS) simulations \cite{pre2005}. This method samples the QS
probability distribution, i.e., the probability distribution at long times,
conditioned on survival; for details on the method see \cite{pre2005}.
The number of realizations used to obtain $m$ varies from $48$ (smallest
size) to $16$ (largest size). Since the particle density can only be varied in
steps of $1 / L^2$, estimates for properties at intermediate values of $p$ are
obtained via interpolation. Figure~\ref{fig:m} shows $m(p)$ in the immediate
vicinity of $p_c$ for $L = 256$ - $2048$. The critical point $p_c$ is
characterized by a finite limiting value, $m_c$ \cite{moments}, so that $p$
values for which $m$ appears to grow or decrease without limit (corresponding to
the sub- and supercritical regimes, respectively), can be excluded as being
off-critical. On the basis of these data, we estimate $p_c = 0.7112687(2)$.

\subsection{The subdiffusive regime}

Next we turn to an analysis of the diffusion coefficient $D$, defined via the
usual relation:

\begin{equation}
\left<[\Delta x]^2 + [\Delta y]^2 \right> = 4 D\,t
\label{meansquare}
\end{equation}

\noindent For each particle $k$, let $\Delta x_k (t) = h^{+}_{x,k}(t) +
h^{-}_{x,k}(t)$ and $\Delta y_k (t) = h^{+}_{y,k}(t) + h^{-}_{y,k}(t)$ represent
its displacement since time zero in the $x$ and $y$ directions, respectively,
due to toppling events. The counting variables $h^\pm_{i,k}(t)$ denote the
number of hops taken by particle $k$ along direction $i$, in the positive and
negative directions, respectively, up to time $t$. The angular brackets in
Eq.~(\ref{meansquare}) denote an average over particles and over independent
realizations of the process. Note that $D$ as defined above is, in general, a
function of time, though we expect it to attain a stationary value at long
times.

Since the displacements of a given particle at successive toppling events are
independent, we expect the msd to grow linearly with the number of topplings.
Although not all displacement attempts are successful (the height restriction
causes some to be rejected), in the stationary regime the fraction of successful
attempts $\eta_s$ is time-independent, as is the density $\rho$ of active sites.
(By $\eta_s (t)$ we mean the number of particles that change their position
during the interval $[t, t+\Delta t]$ divided by twice the number of toppling
events during this interval.) Thus the msd should grow linearly with {\it time}
in the stationary regime, as is verified here and in the one-dimensional case
\cite{epjb2009}. By the same reasoning, if $\eta_s$ and/or $\rho$ vary with
time, we should expect deviations from the linear relation $\langle [\Delta x]^2
+ [\Delta y]^2 \rangle \propto t$.

For $p$ near $p_c$, the initial activity density generated by random particle
insertion is much larger than the stationary activity density. This is easily
seen by noting that, during the insertion process, setting the rate of particle
insertion {\it attempts} at one particle per site and per unit time, $s$ the
fractions $f_j (s)$ of sites bearing exactly $j$ particles satisfy the
equations~\footnote{To avoid confusion, we denote the time variable associated
with
the initial filling process by $s$. When the particle density has attained its
desired value $p$, $s = s_f$}:

\begin{equation}
\frac{df_0}{ds} = -f_0, \;\;\;\;\;\;
\frac{df_1}{ds} = f_0 - f_1, \;\;\;\;\;\; \mbox{and} \;\;\;
\frac{df_2}{ds} = f_1.
\label{fjt}
\end{equation}

Since the lattice is empty at time zero we have, $f_0 = e^{-s}$, $f_1 = s
e^{-s}$, and $f_2 = 1 - (1 + s)e^{-s}$. The process stops when the particle
density $f_1 + 2f_2$ reaches the desired value $p$. (Thus the stopping time
$s_f$ is related to $p$ via a transcendental equation.) For particle density
$p_c \simeq 0.71127$, this yields an initial activity density of $f_2 \simeq
0.1778$, much larger than the stationary activity density, which in fact tends
to zero as $L \to \infty$ at $p = p_c$. Thus we should expect the activity
density $\rho(t)$ to decrease initially.

\begin{figure}[htb]
\begin{center}
\vspace{0.5cm}
(a)~~~~\includegraphics[scale=0.29]{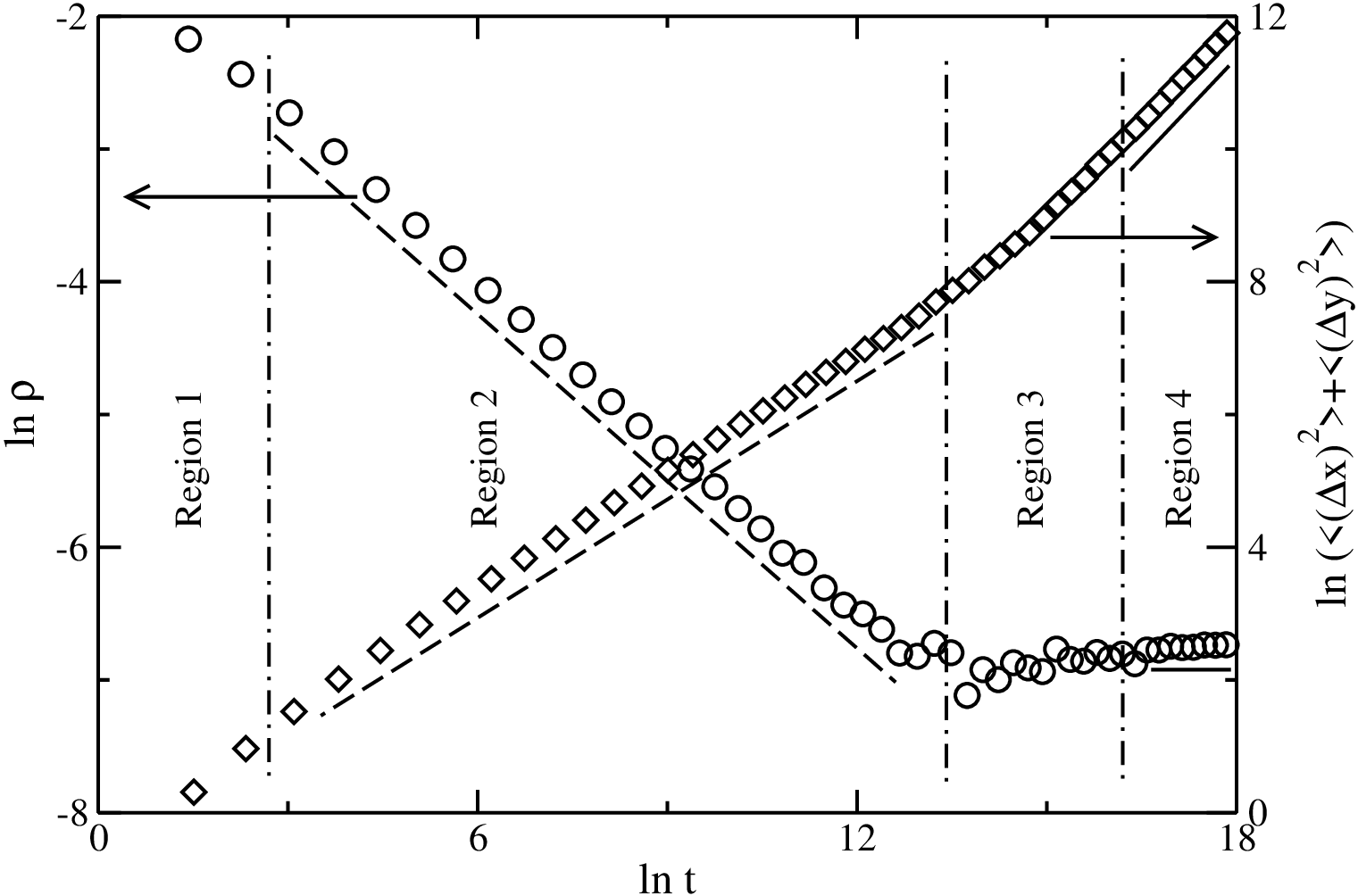}~~~
\includegraphics[scale=0.29]{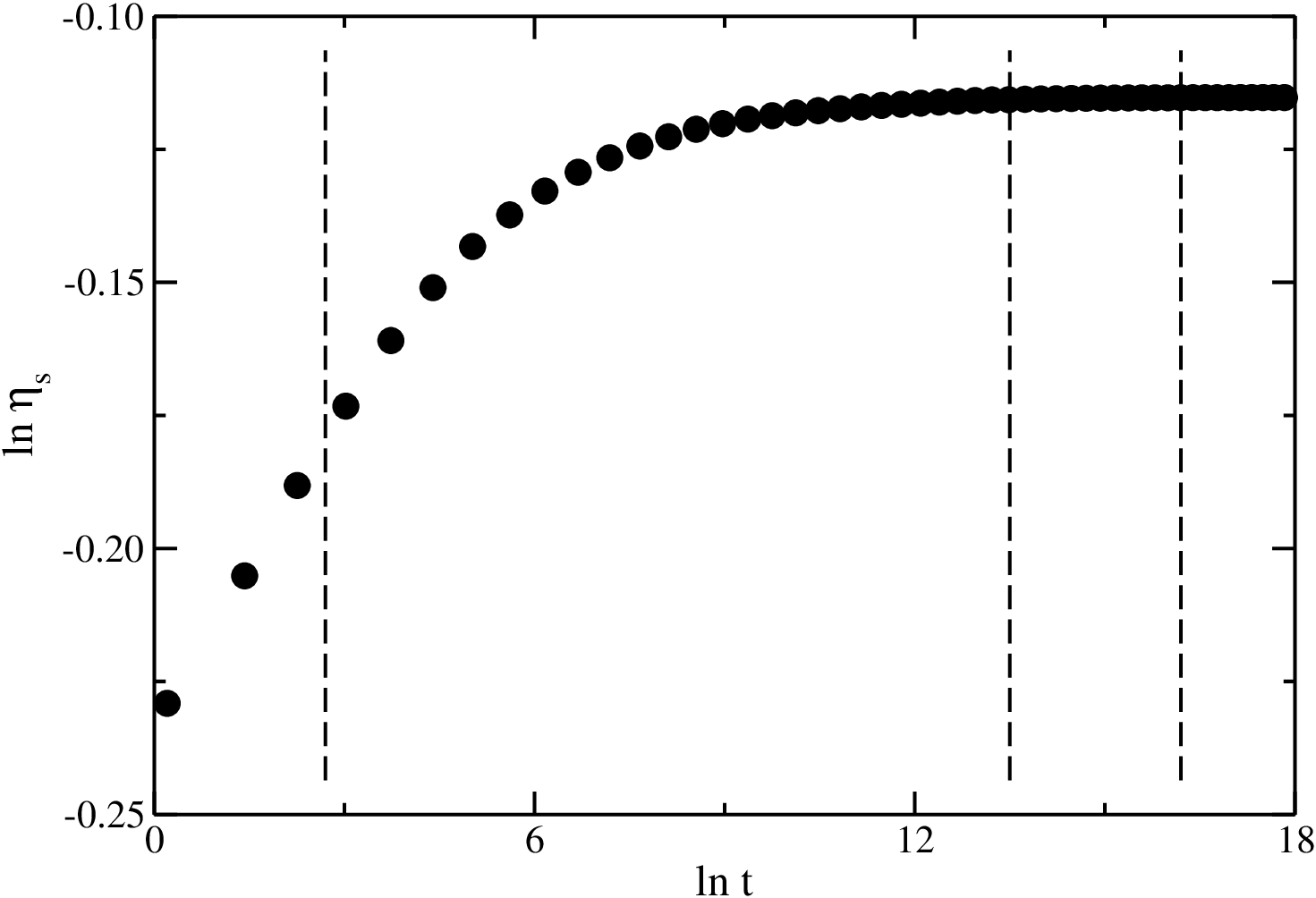}~~(b)
\caption{\label{fig:msd_rho} (a) msd and $\rho$ versus time for system size
$L=2048$ and particle density $p = 0.711315$, slight above $p_c$. The solid
lines correspond the stationary regime (region ``4''), broken lines to the
subdiffusive regime (region ``2''). (b) Fraction of successful transfer
attempts $\eta_s$ versus time $t$.\\}
\end{center}
\end{figure}

In fact, we can divide the sandpile evolution into four regimes, as shown in
Fig.~\ref{fig:msd_rho} (a): (1) an initial transient; (2) the subdiffusive
regime, in which the msd grows more slowly than linearly while $\rho$ decays
as a power law; (3) a regime of anomalous growth in the activity $\rho(t)$;
(4) the stationary regime. Analogous time evolution of the activity is reported
in simulations of a Langevin equation for the two-dimensional CSS in
Ref.~\cite{ramasco}. In the the subdiffusive we find $\langle [\Delta x]^2 +
[\Delta y]^2 \rangle \propto t^\gamma$, with $\gamma <1$, as determined via a
least-squares linear fit to the msd data on log scales. Figure~\ref{fig:msd_rho}
(b) shows that the fraction $\eta_s$ of successful particle displacement
attempts increases slowly with time in the subdiffusive regime, but not enough
to compensate for the sharp reduction in $\rho$. The steady decrease in
$\rho(t)$ therefore generates sublinear growth in the msd.

\begin{figure}[htb]
\begin{center}
\vspace{0.5cm}
(a)~~~~\includegraphics[scale=0.29]{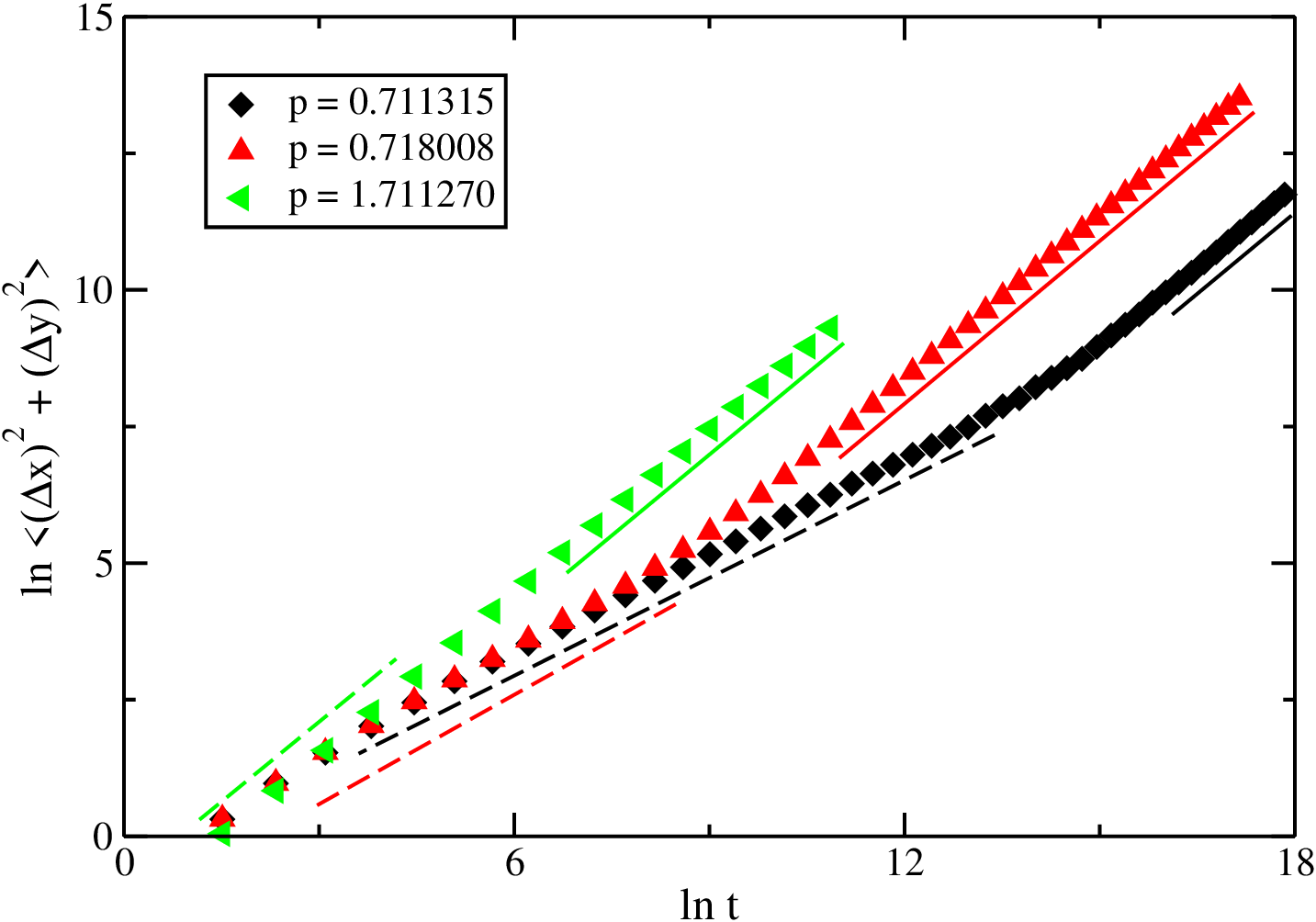}~~~
\includegraphics[scale=0.29]{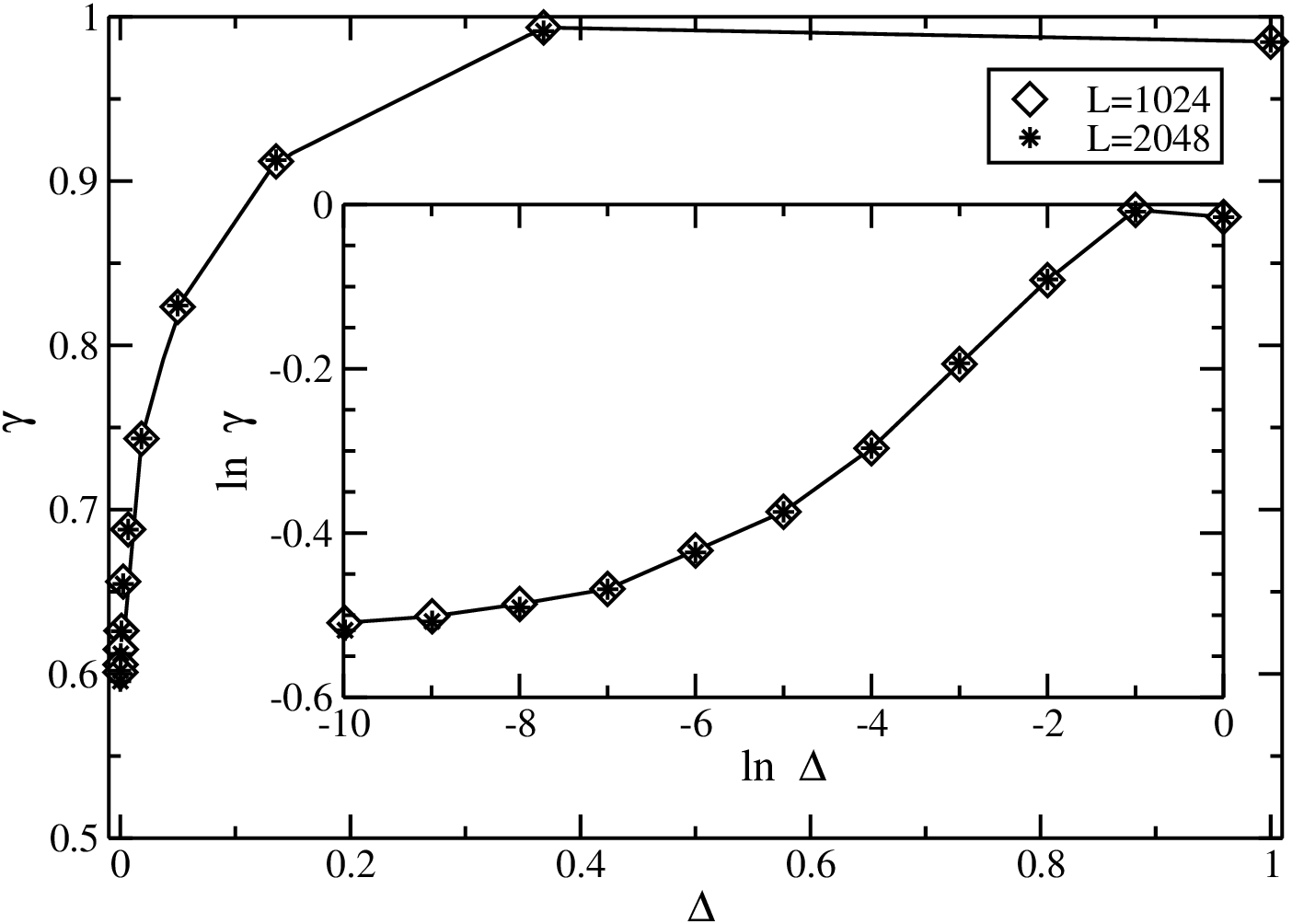}~~(b)
\caption{\label{fig:msd} (Color online) (a) msd versus time for $L = 2048$ and p
values as indicated. The solid lines have a slope of unity; the slopes of the
broken lines $\gamma$ vary from $\approx$ $0.60$ to $\approx$ $0.98$. (b)
$\gamma$ versus $\Delta = p - p_c$ in the subdiffusive regime. Inset: the same
data on logarithmic scales.\\}
\end{center}
\end{figure}

\begin{figure}[htb]
\begin{center}
\vspace{0.5cm}
\includegraphics[scale=0.29]{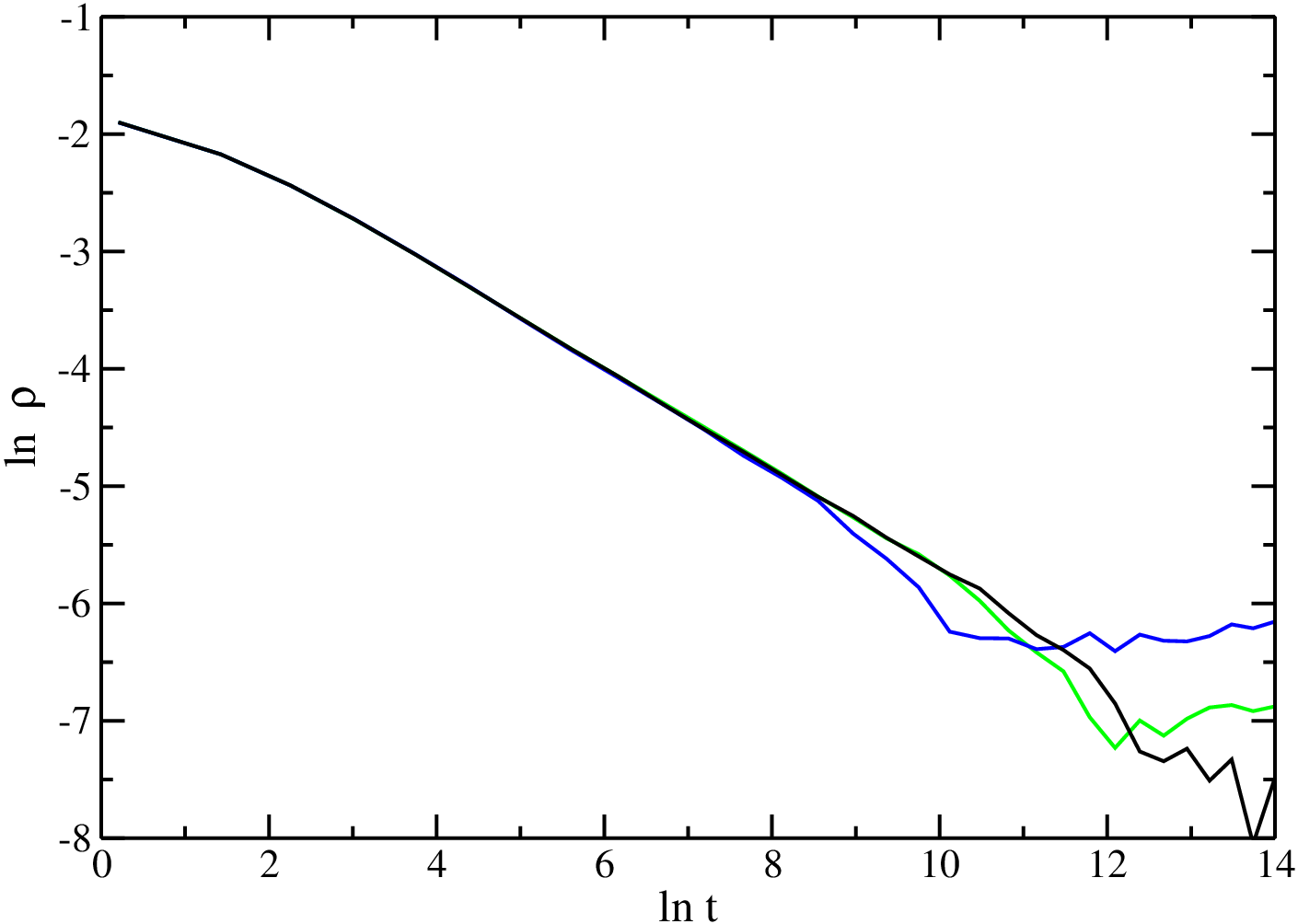}
\caption{\label{fig:rvt} Activity density $\rho$ versus time $t$ at $p\!=\!p_c$,
for system sizes (upper to lower at right) $L=512$, $1024$, and $2048$.}
\end{center}
\end{figure}

The msd, $\left<[\Delta x]^2 + [\Delta y]^2\right>$ is plotted versus time in
Fig.~\ref{fig:msd} (a) for three values of $p$ in the supercritical regime; the
crossover between the short-time subdiffusive regime and the long-time
linear regime is evident. Figure \ref{fig:msd} (b) shows the dependence
of $\gamma$ on $\Delta \equiv p - p_c$ in the subdiffusive regime. We note that,
away from criticality, $\gamma$ is not expected to represent a universal scaling
property. At the critical point, we expect the power-law decay $\rho(t) \sim
t^{-\delta} $ in the initial (``short-time") scaling regime. We verify this
scaling law at criticality, in the subdiffusive regime in Fig.~\ref{fig:rvt}.
Our results yield $\delta = 0.436(2)$, $0.439(3)$, and $0.427(2)$ for system
sizes $L=512$, 1024, and 2048, respectively; we adopt $\delta = 0.43(1)$ as our
best estimate. For system size $L=2048$ we find $\gamma = 0.585(1)$ at
criticality. We expect each particle to execute a random walk in which the
number of steps per unit time is proportional to the activity density $\rho$.
Thus, in the subdiffusive regime, particle diffusion can be
described qualitatively by a Langevin equation,

\begin{equation}
\frac{d\, {\bf x}}{dt} = {\bf A(t)}
\label{langevin}
\end{equation}

\noindent in which the noise satisfies $\langle A_i(t) A_j(s)\rangle = \Gamma
\delta_{ij} \rho(t) \delta(t-s)$ (here $\Gamma$ is a constant and $i$ and $j$
are Cartesian indices). Integrating the Langevin equation one readily
finds that the msd grows $\propto t^{1-\delta}$, implying that $\gamma + \delta
= 1$. Our numerical results are essentially consistent with this, yielding
$\gamma + \delta = 1.015(11)$.

\subsection{Static scaling behavior}

\begin{figure}[htb]
\begin{center}
\vspace{0.5cm}
(a)~~~~\includegraphics[scale=0.29]{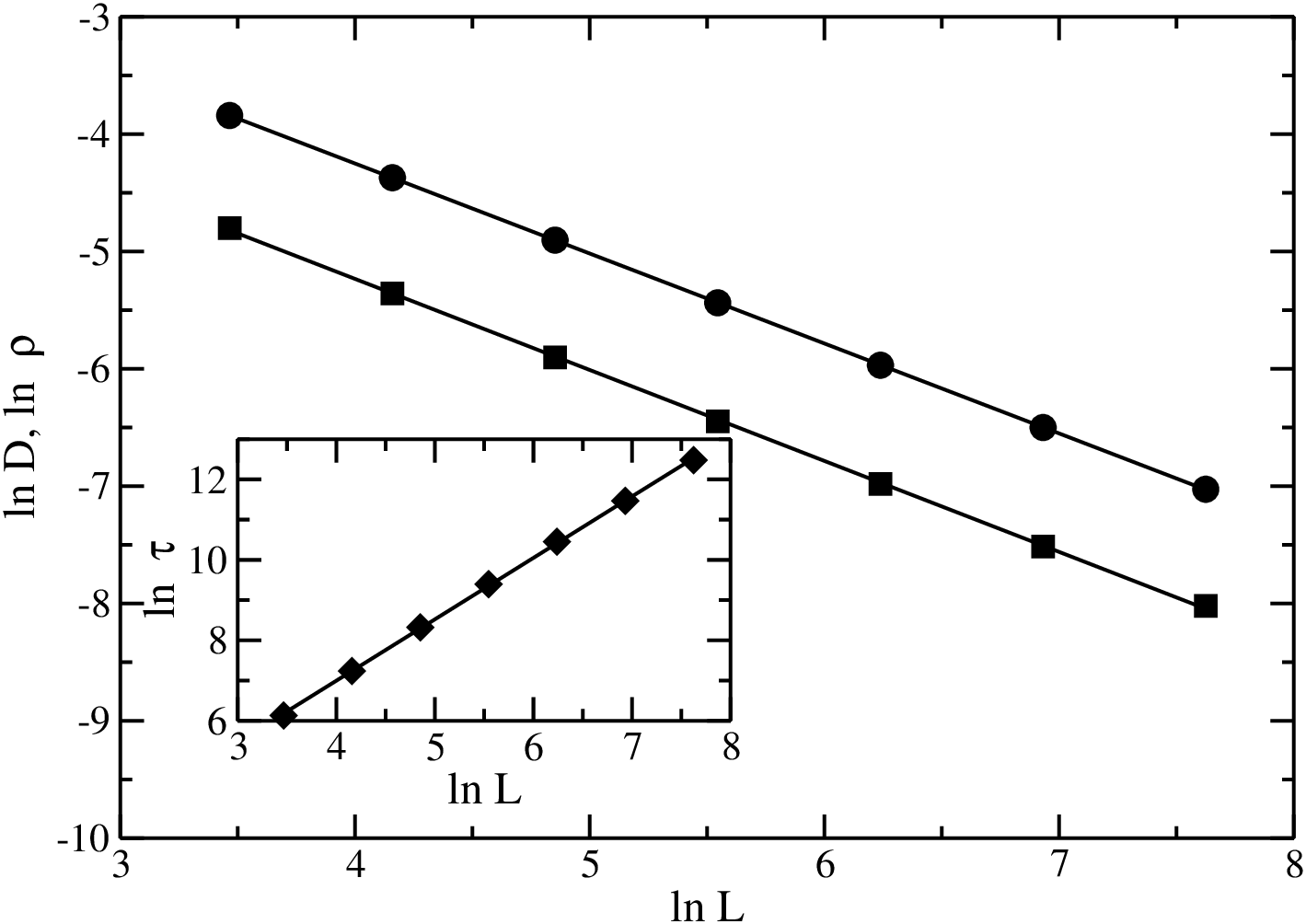}~~~
\includegraphics[scale=0.29]{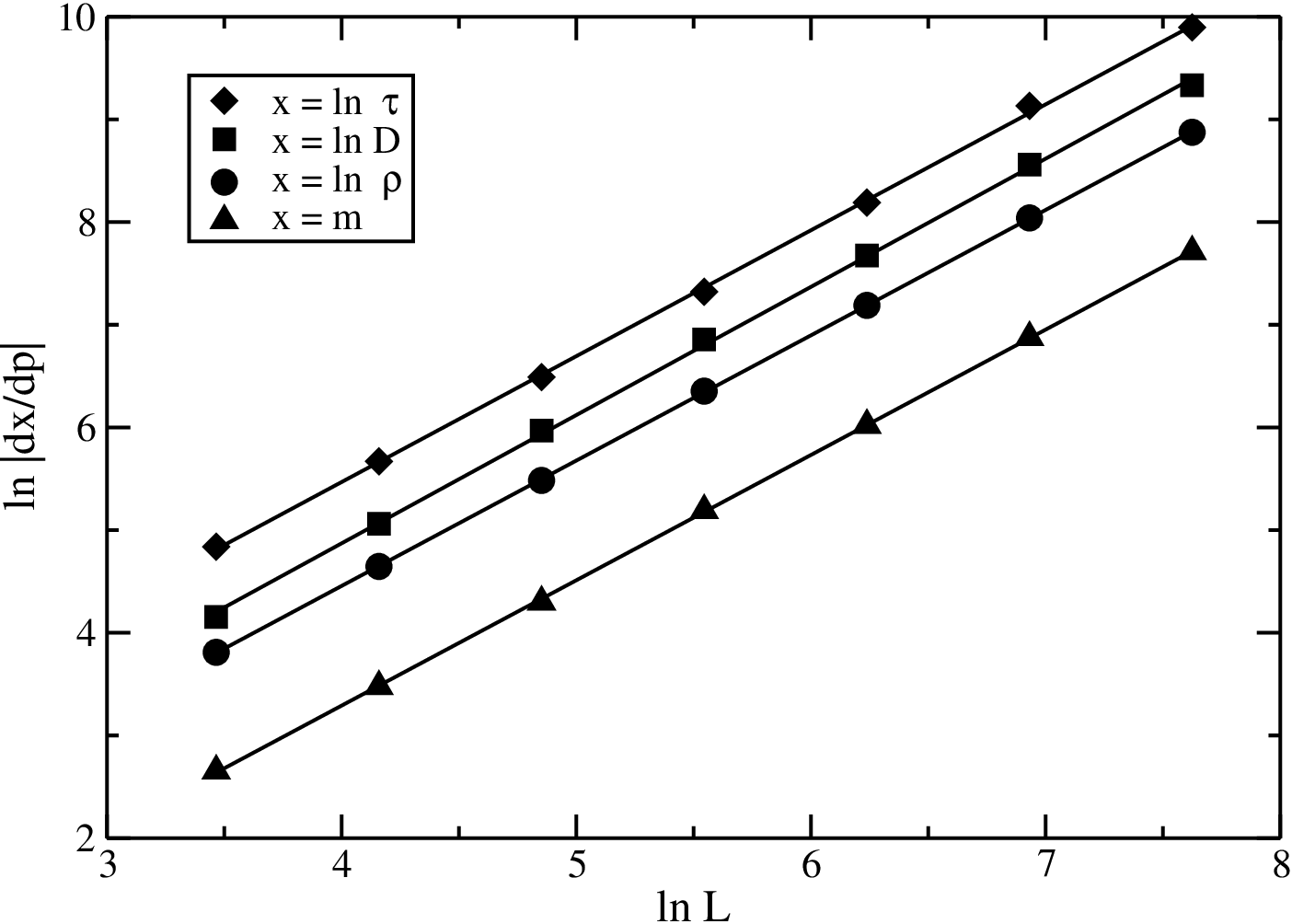}~~(b)
\caption{\label{fig:critical_point_derived} (a) Diffusion constant $D$
(squares) and stationary order parameter $\rho$ (circles) versus system size $L$
at the critical point. The slopes of the straight lines are (lower to upper)
$0.768(6)$ and $0.779(6)$. Inset: lifetime $\tau $ (diamonds) versus $L$; the
slope of the straight line is $1.519(8)$. (b) Derivatives of (lower to upper)
$m$, $\ln \rho$, $\ln D$ and $\ln \tau$ with respect to particle density,
evaluated at $p_c$, versus system size $L$. The slopes of straight lines are
$\approx$ $1.234(8)$.}
\end{center}
\end{figure}

Next we analyze the stationary values of $\rho$, $D$ and $\tau$ (the
mean lifetime) in the critical region. For the two first parameters, we use QS
simulations. The number of realizations varies from $1024$ (for $L = 32$) to
$16$ (for $L = 2048$). To begin, we estimate $\rho$ and $D$ at the critical
point; $\rho$ is expected to scale with system size as $\rho \sim
L^{-\beta/\nu_\bot}$. Our results permit us to estimate critical exponent ratio,
and confirm that $D$ and $\rho$ scale in the same manner, as can be seen in
Fig.~\ref{fig:critical_point_derived} (a).

The mean lifetime $\tau$ is determined from the probability $P(t)$ that the
activity survives up time $t$. We used conventional (not QS) simulations in this
case. The number $N_s$ of independent realizations used in this set varies from
$65536$ (for the smallest size) to $1024$ (for the largest). We verify that
$P(t)$ decays exponentially, so that the mean lifetime can be defined via
$P(t) \sim \exp(-t/\tau)$. At the critical point, standard finite-size scaling
arguments yield $\tau \sim L^{z}$, with $z$ the dynamical exponent, given by the
scaling relation $z=\nu_{||}/\nu_\bot$. Discarding the initial transient of each
study,the survival time $\tau$ is estimated by fitting of the exponential tail
of $P(t)$, as well as by determining the time required for the survival
probability to decay to one half \cite{pre2000}. Since the particle density can
only be varied in steps of $1 / L^2$, estimates for properties at intermediate
values of $p$ are obtained via interpolation. Figure
\ref{fig:critical_point_derived} (a) shows the behavior of $\rho$, $D$ and
$\tau$ at the critical point. Using least-squares linear fits to the data for
$\rho$, $D$ and $\tau$, we obtain the estimates $\beta/\nu_\bot=0.768(8)$,
$\beta/\nu_\bot=0.779(6)$ and $z=1.519(8)$, respectively.

We determine the correlation length exponent $\nu_\bot$ in the following manner.
In the immediate vicinity of critical point, the derivatives of the moment ratio and
of $\ln \rho$ with respect to $p$ scale as \cite{pre2000}:
\begin{equation}
\left|\frac{\partial \, m}{\partial p}\right|_{p_c} \propto
L^{1/\nu_{\bot}}
\label{derm}
\end{equation}
and
\begin{equation}
\left|\frac{\partial \ln \rho}{\partial p}\right|_{p_c} \propto
L^{1/\nu_{\bot}}
\label{deffD}
\end{equation}
\noindent Similar behaviors hold for the derivatives of $\ln D$ and $\ln \tau$. (The
derivatives are estimated numerically using the central difference method, using
the values of $\ln \rho$, $\ln D$ and $\ln \tau$ above, below and at $p_c$.)
Figure \ref{fig:critical_point_derived} (b) shows the derivatives as a function
of system size. Least-squares linear fits to the data yield $\nu_\bot =
0.818(6)$, $\nu_\bot = 0.817(6)$, $\nu_\bot =0.795(6)$ and $\nu_\bot
= 0.811(6)$ for $m$, $\ln \rho$, $\ln D$ and $\ln \tau$, respectively, for a
global estimate of $\nu_\bot \approx 0.812(7)$. Then, using $\beta/\nu_\bot
=0.779(6)$ obtained above, we find $\beta = 0.633(6)$. Similarly, using
$z=\nu_{||}/\nu_\bot =1.512(8)$, we find $\nu_{||} = 1.225(7)$.

\subsection{Supercritical regime}

\begin{figure}[htb]
\begin{center}
(a)~~~~~\includegraphics[scale=0.29]{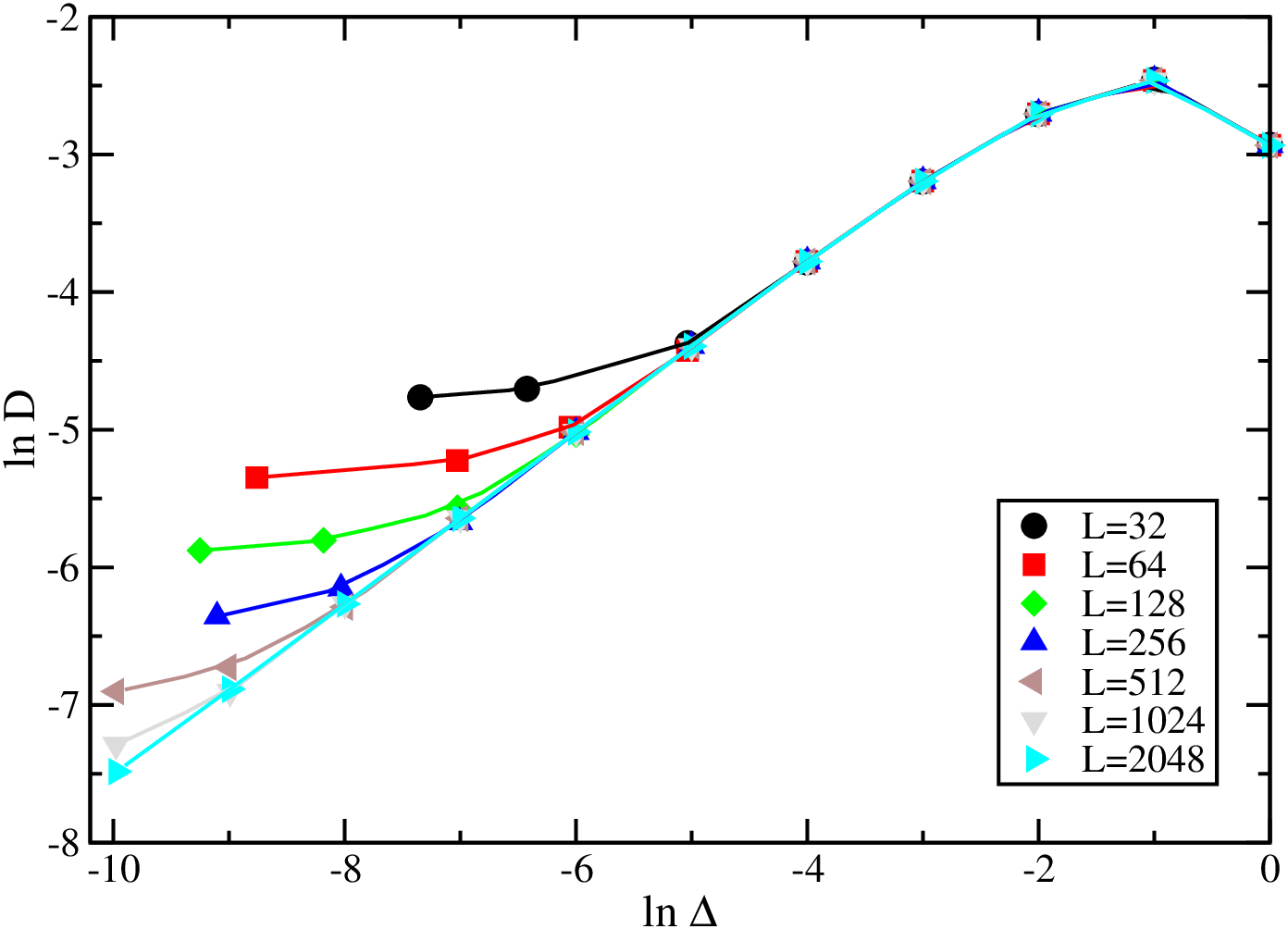}~~~
\includegraphics[scale=0.29]{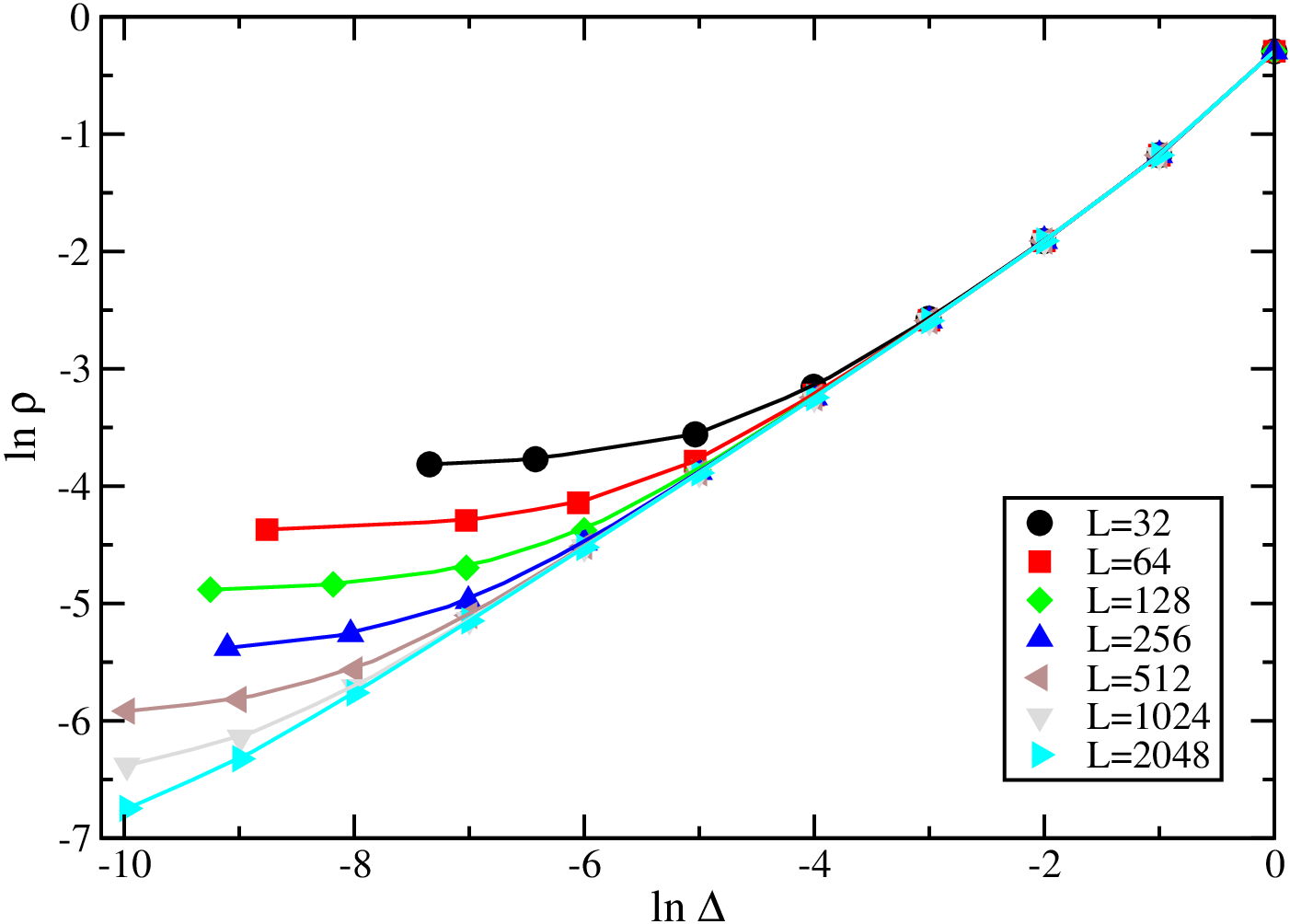}~~(b)
\caption{\label{fig:lnD_lnrho} (Color online) (a) Diffusion rate versus $\Delta$
for system sizes as indicated. Error bars are smaller than symbols.
(b) Stationary activity density $\rho$ versus $\Delta$.}
\end{center}
\end{figure}

\begin{figure}[htb]
\begin{center}
\includegraphics[scale=0.29]{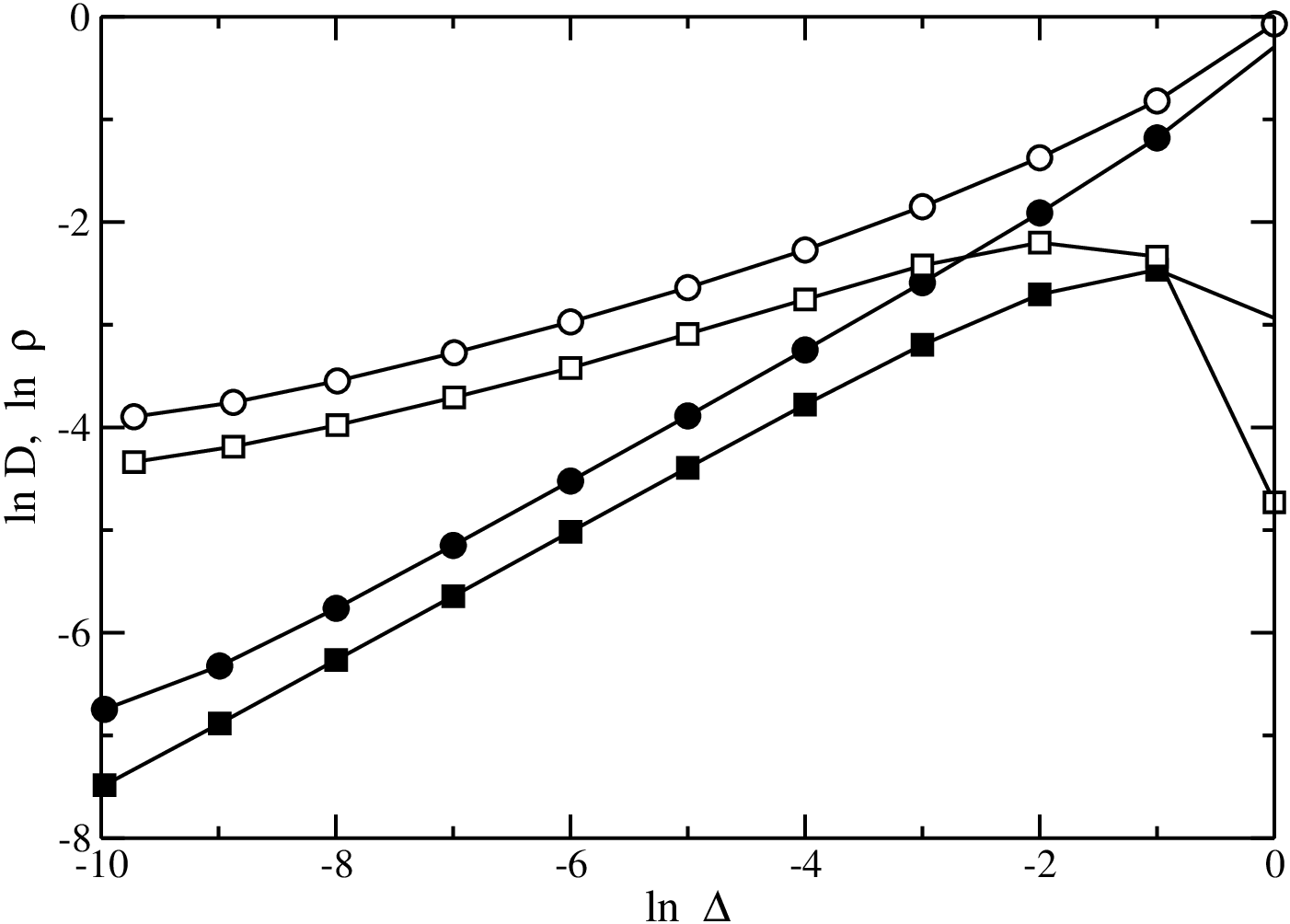}
\caption{\label{fig:compare} Stationary activity (circles) and asymptotic
diffusion rate (squares) versus $\Delta$ in one dimension (open symbols),
system size $L=50000$, and two dimensions (filled symbols), system size
$L=2048$.}
\end{center}
\end{figure}

We determined the stationary activity density and diffusion rate for a series of
particle densities $p$ in the supercritical regime, using QS simulations.
In this set, the number of realizations varies from $512$ (smallest
size) to $8$ (largest size) for values of $p$ near $p_c$, and $3$ for values
distant from $p_c$. The time required for the diffusion constant to reach its
stationary value is somewhat greater than for the order parameter, e.g., $t_D  =
3 \times 10^7$ and $t_\rho = 2 \times 10^7$ for the largest system studied and
$p = 0.7112692$. Figure \ref{fig:lnD_lnrho} shows the stationary values of $D$
and $\rho$ versus $\Delta = p - p_c$. Several observations can be made: 1) for
the largest systems studied, $D$ and $\rho$ converge to their limiting ($L
\rightarrow \infty$) values for $\Delta \geq 0.0025$ and $\Delta \geq 0.0015$
respectively; 2) even for values of $\Delta$ such that the diffusion rate has
converged, the slope of $D(\Delta$) on logarithmic scales changes appreciably
with $\Delta$, making a reliable estimate of the critical exponent $\beta$
difficult, using these data; and 3) $D$ drops sharply when $p \rightarrow 2 $,
due to the height-restriction, which inhibits the particle motion. These
tendencies are also observed in the one-dimensional model, as can be seen
in Fig.~\ref{fig:compare}. Since the curvature of ln$\rho$ and ln$D$ as
functions of ln$\Delta$ is less pronounced in the two-dimensional case, we
performed linear fits to the data for $\rho$ and $D$ on intervals for which the
curvature of the log-log plot is minimal (that is, for $-8 \leq \ln \Delta \leq
-2$, and $-10 \leq \ln \Delta \leq -3$, respectively).  The resulting estimates
for the critical exponent $\beta$ are $0.64(1)$ (from the activity density data)
and $0.62(1)$ (from the diffusion coefficient); these values are in reasonable
agreement with the estimate obtained above via scaling analysis.\\

\begin{figure}[htb]
\begin{center}
(a)~~~~\includegraphics[scale=0.29]{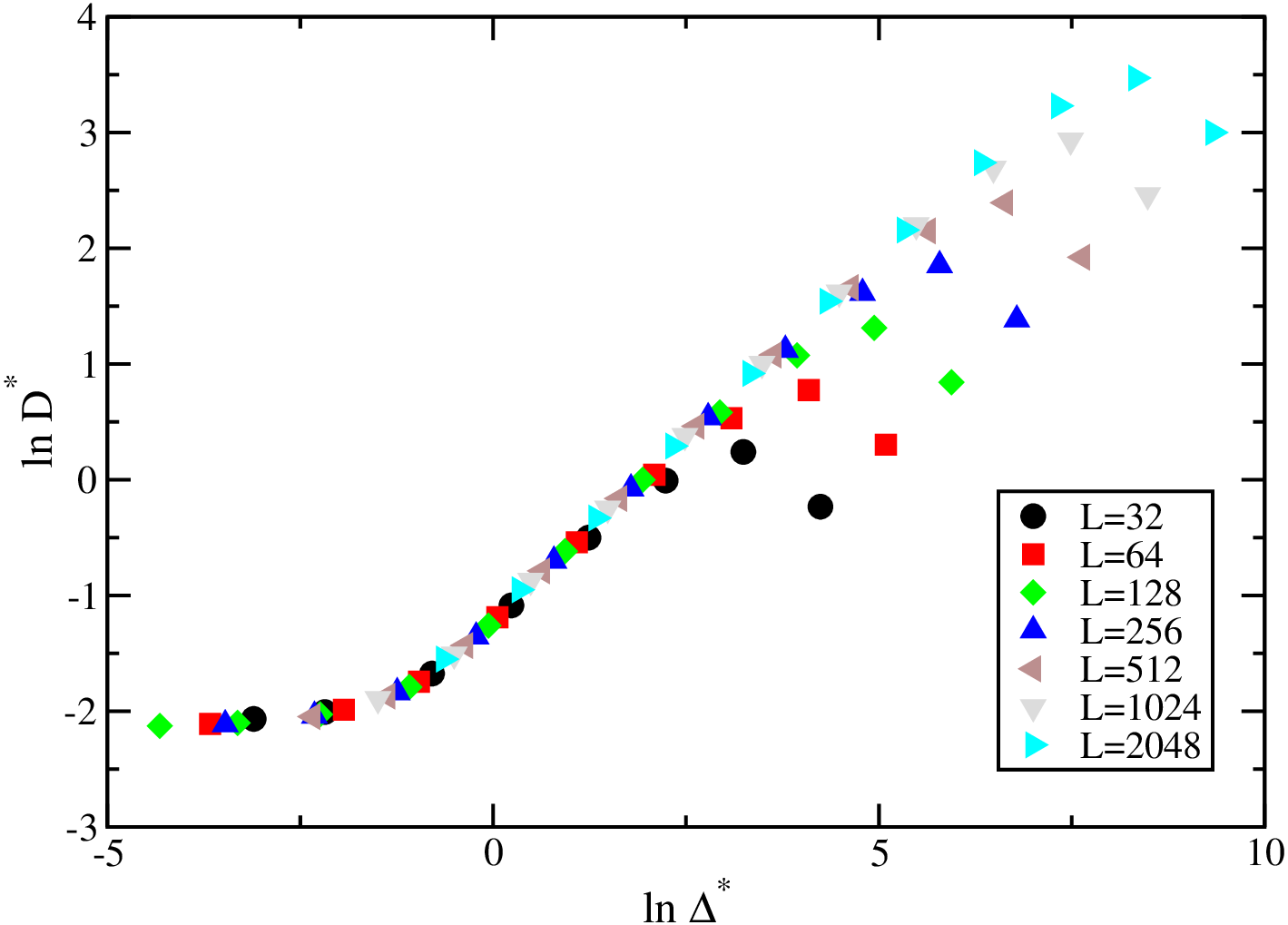}~~~
\includegraphics[scale=0.29]{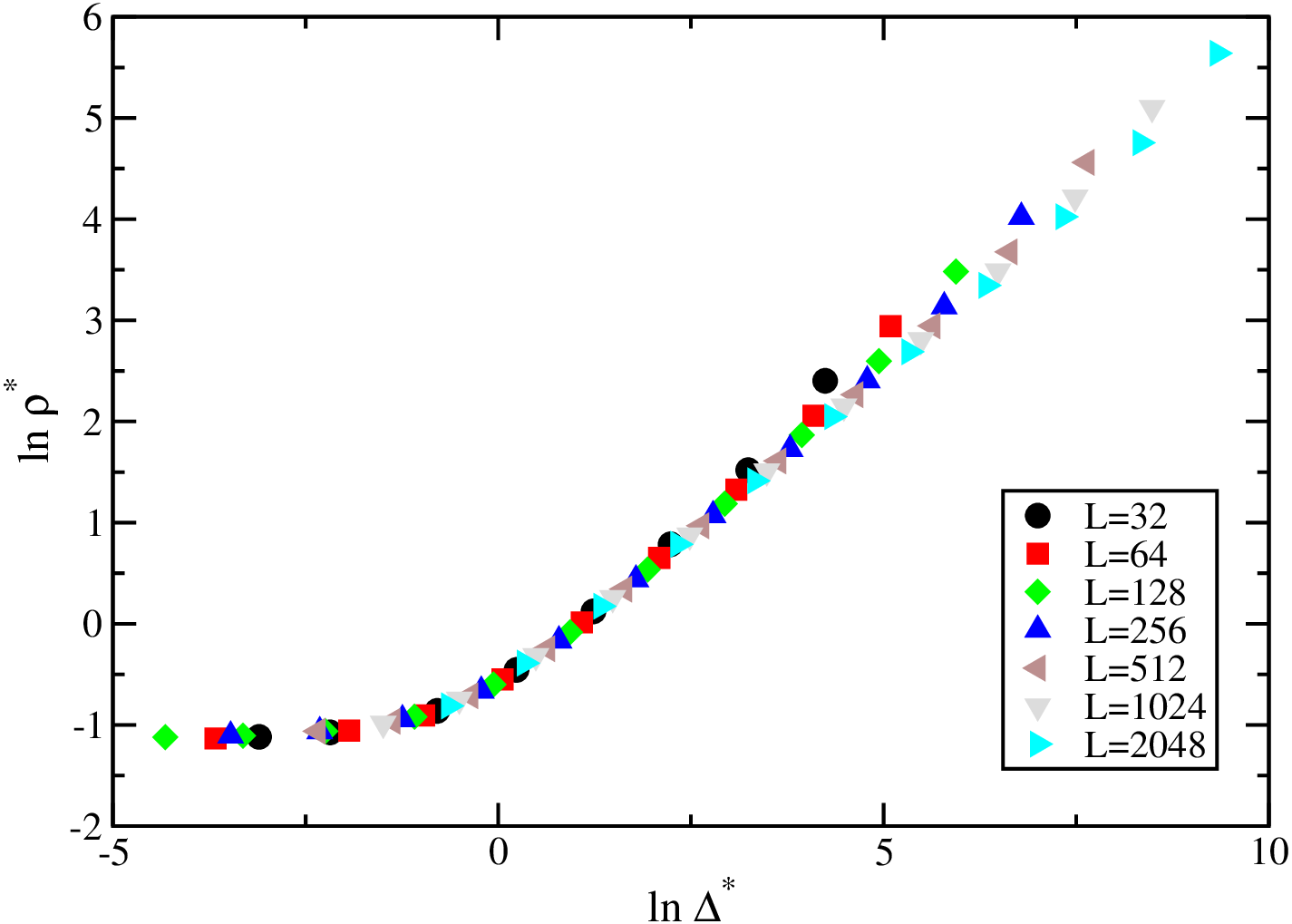}~~(b)
\caption{\label{fig:D_rho} (Color online) (a) Scaled diffusion rate $D^*$ versus
scaled distance from critical point $\Delta^*$, as defined in text. (b) Scaled
activity $\rho^*$ versus $\Delta^*$. }
\end{center}
\end{figure}

From Fig.~\ref{fig:lnD_lnrho}, it is clear that neither $D(\Delta)$ nor
$\rho(\Delta)$ have a simple power-law dependence on $\Delta$, even for
parameter values such that there is no discernible finite-size effect for $D$
and $\rho$. In the case of the diffusion constant, the data collapse for $\Delta
\geq 0.018$, for all sizes studied, and $\Delta \geq 0.00013$, for $L = 1024$
and $L = 2048$. The activity exhibits a rather similar behavior when $\Delta
\rightarrow 0$. In Fig.~\ref{fig:D_rho} we test whether the data for the scaled
activity, $\rho^{*} = L^{\beta/\nu_{\bot}}\rho$, collapse when plotted versus
the scaled distance from criticality, $\Delta^{*}=L^{1/\nu_{\bot}}\Delta$. Data
collapse is observed only quite near the critical point, consistent with
previous studies of the CSS \cite{pre2006,epjb2009}. $D$ and $\rho$ collapse
with the same critical parameters. The critical exponents obtained via data
collapse are $\beta = 0.64(1)$ and $\nu_\bot = 0.84(1)$. These results are
consistent with the critical exponents found using our first set of simulations, and
with the best numerical estimates obtained from simulations of the
Langevin equation: $\beta = 0.66(2)$, $\nu_\bot = 0.84(2)$, and $\nu_{||} =
1.27(7)$ \cite{ramasco}.\\

\subsection{Activity correlation function}

\begin{figure}[htb]
\begin{center}
\vspace{0.5cm}
(a)~~~~\includegraphics[scale=0.29]{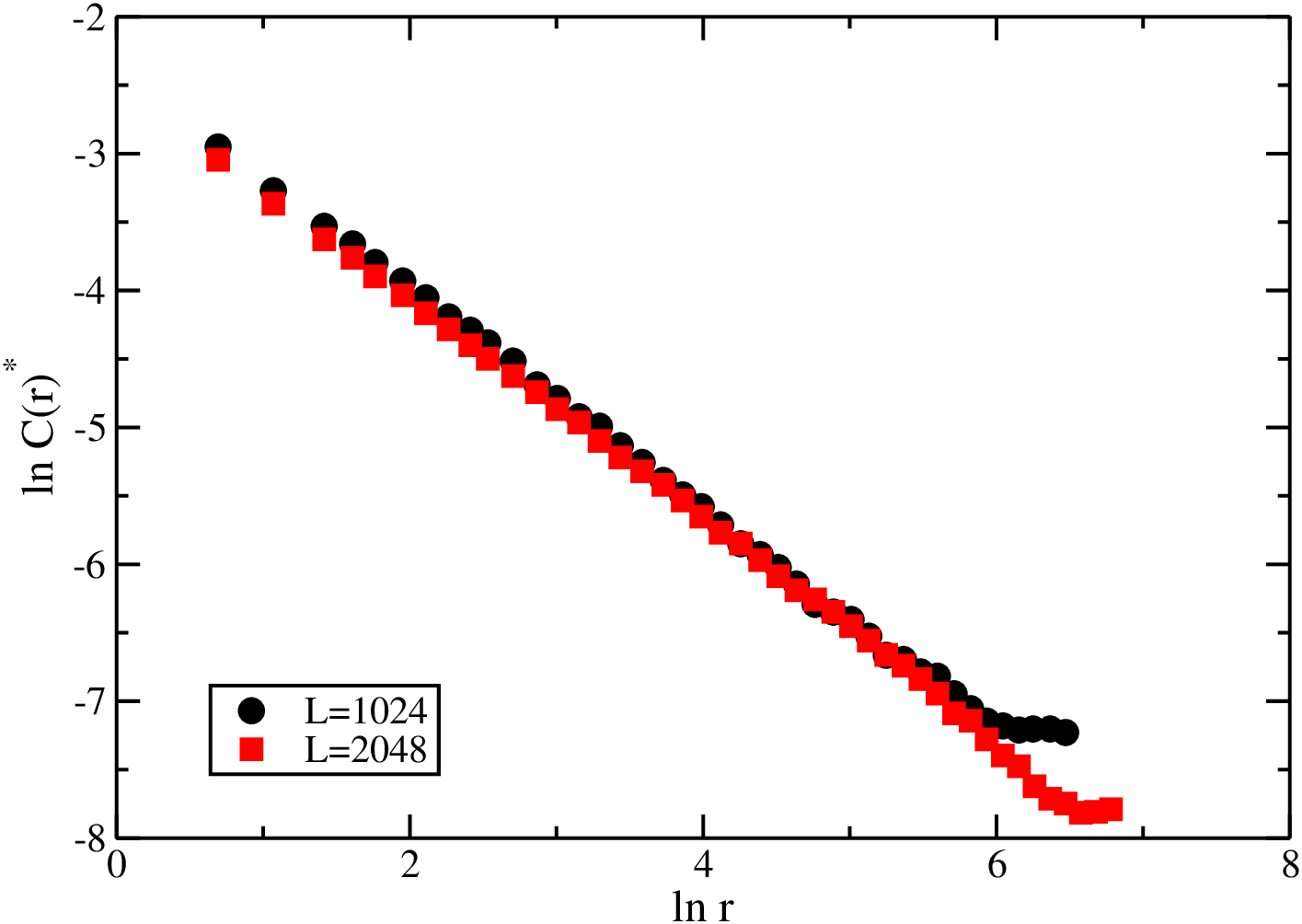}~~~
\includegraphics[scale=0.29]{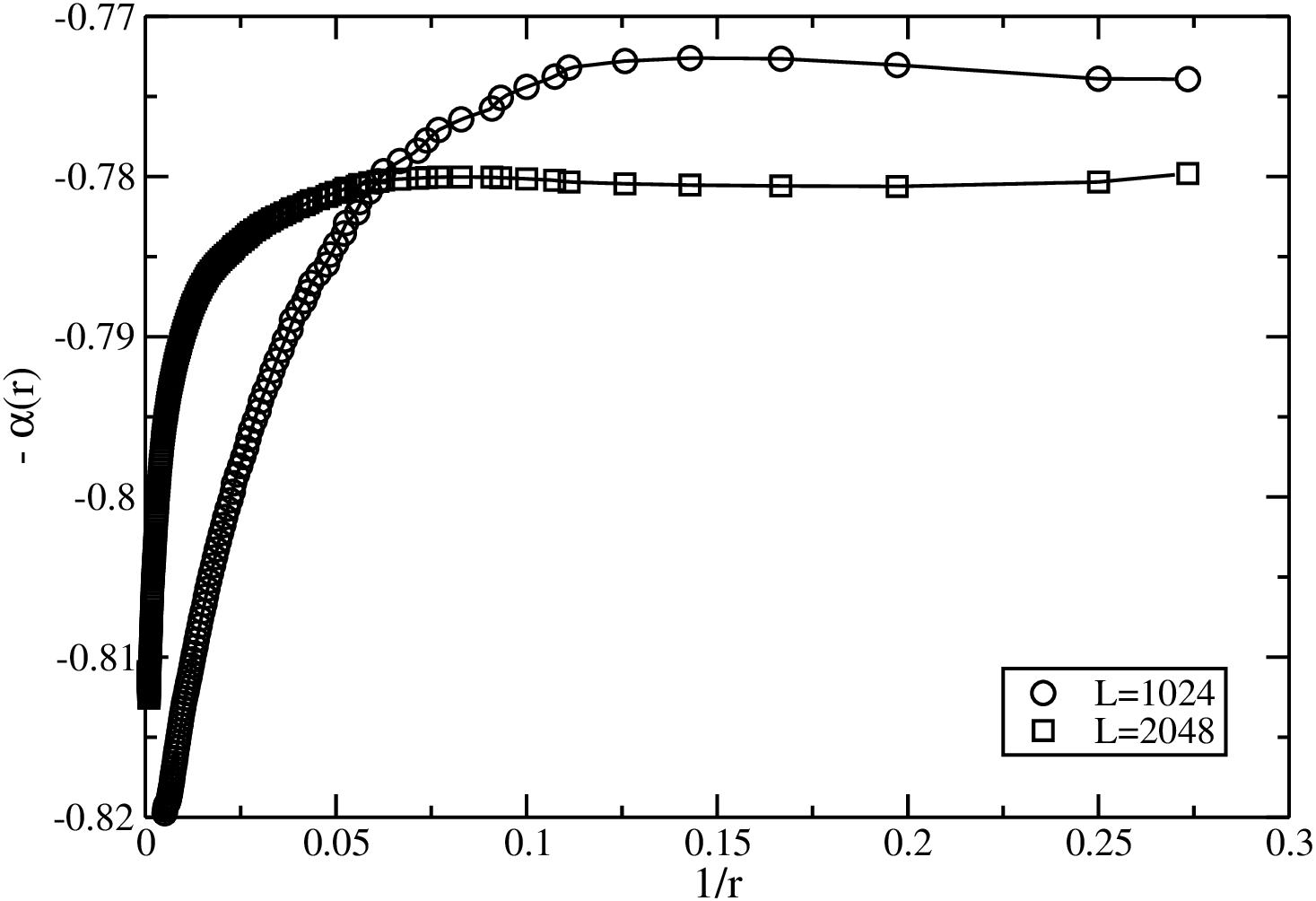}~~(b)
\caption{\label{fig:correlation} (a) The Static correlation function
modified $C^*(r)$ versus r for $L = 1024$ and $L=2048$. (b) Estimate for the exponent
$\alpha = \beta/\nu_{\bot}$ versus $1/r_f$, as defined in text.}
\end{center}
\end{figure}

To our knowledge, few if any studies have been performed on spatial correlations
of the activity in stochastic sandpiles.  Here we use QS simulations to
determine the static correlation function, defined via
\begin{equation}
C(r) = \left<A_{i,j} A_{k,l} \right> -
\left<A_{i,j}\right>\left<A_{k,l}\right>
\label{corr1}
\end{equation}

\noindent where
$r = \sqrt{(i-k)^2+(j-k)^2}$ and $A_{i,j} \equiv 1$ if $z_{i,j} = 2$, and is
zero otherwise. (That is, $A_{i,j}$ is the indicator variable for activity at
the site.) Note that $\langle A_{i,j} \rangle = \rho$. At the critical point of
an absorbing-state phase transition, correlations are expected to decay as a
power law, $C(r) \propto r^{-\beta/\nu_{\bot}}$, for $r \ll L$ \cite{qsscp}. In
Fig.~\ref{fig:correlation} (a) we plot $C^*(r) = L^{\, \beta/\nu_{\bot}} C(r)$
for $L = 1024$ and $L=2048$, using the estimate for $\beta/\nu_{\bot}$ obtained
above; evidently the two curves collapse, and follow essentially the same power
law. For $r = L/2$ the correlation function attains a minimum, as expected due
to the periodic boundaries conditions. To obtain the decay exponent we analyze
the local slope $\phi$ (see Fig.~\ref{fig:correlation} (b)), obtained from a
linear fit to the data for ln $C$ versus ln $r$ (the points being equally spaced
in ln$r$), fixing the initial point $r_i = 2$ and varying the final point $r_f$
included in the fitting interval. The resulting values for $\beta/\nu_{\bot}$
yield the estimate $0.777(8)$, consistent with our previous result.
\vspace{0.5cm}

\subsection{Particle mobility}

\begin{figure}[htb]
\begin{center}
\vspace{0.5cm}
(a)~~~~\includegraphics[scale=0.29]{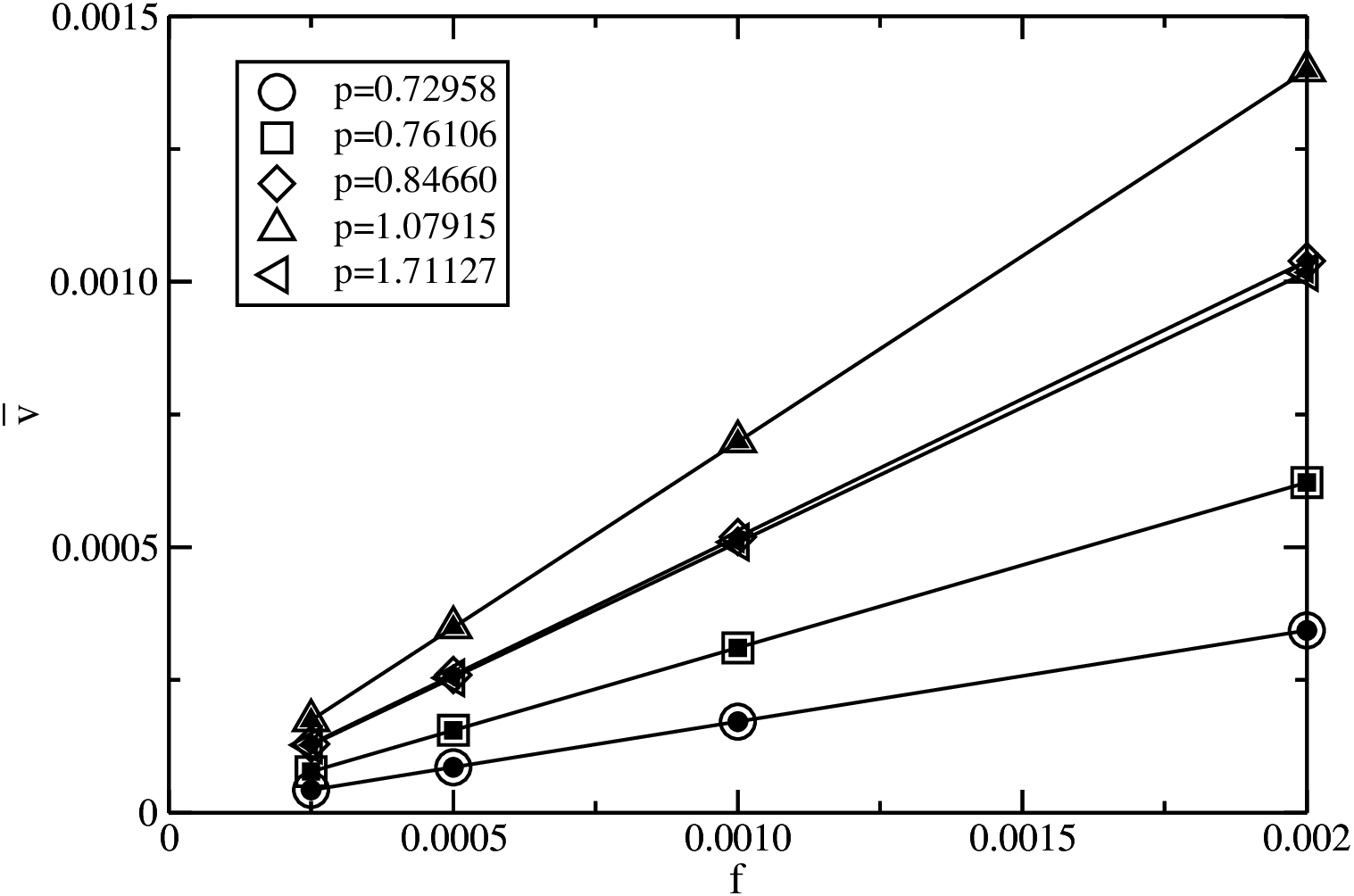}~~~
\includegraphics[scale=0.29]{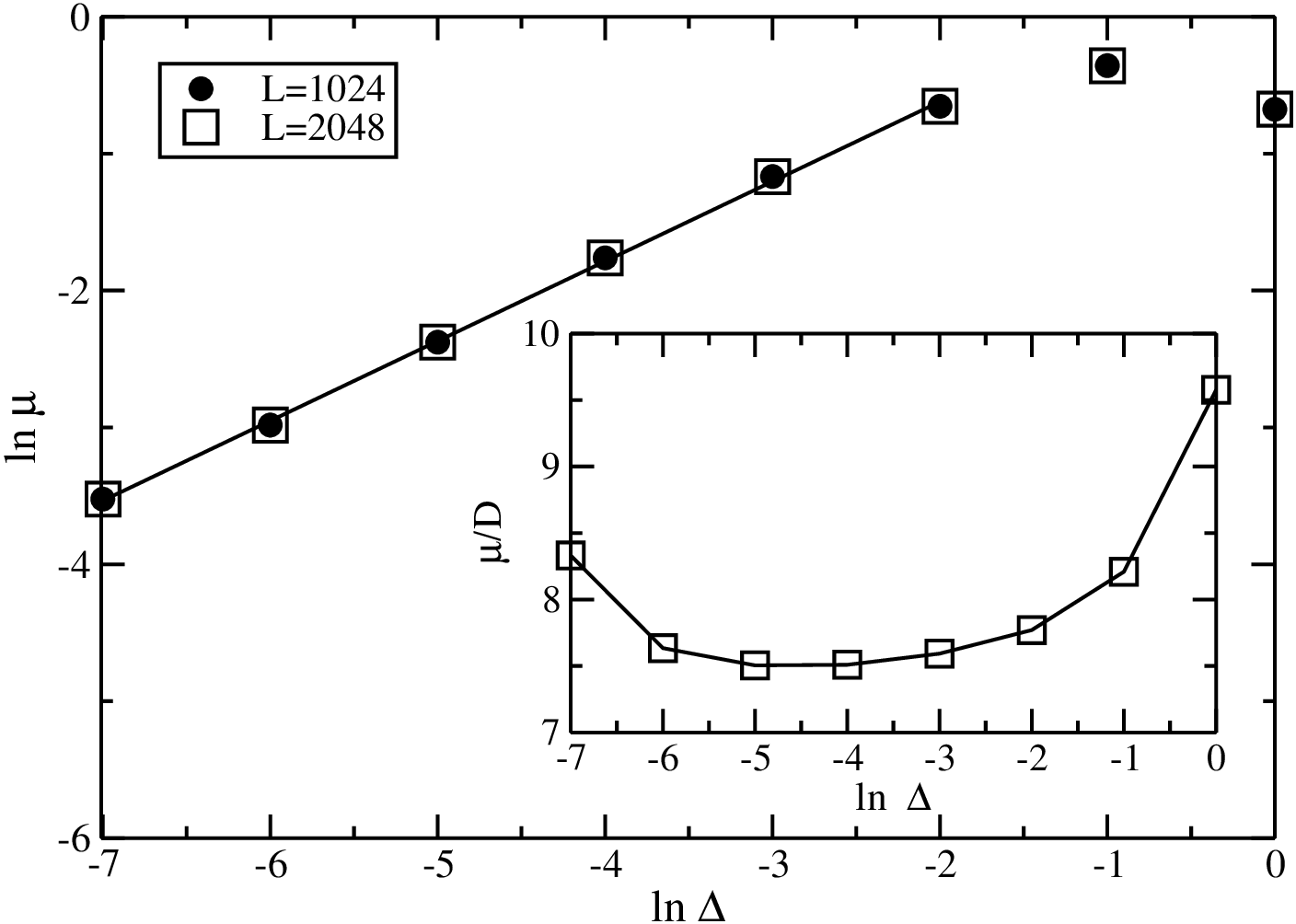}~~(b)
\caption{\label{fig:meanv} (a) Mean rate of particle displacement
$\overline{v}$ versus bias $f$ for particle densities $p$ as indicated and
system sizes $L = 1024$ (filled symbols) and $2048$ (open symbols). (b) $\mu$
versus $\Delta$. The slope of the solid line is $0.5829(1)$. Inset: Stationary
value of $\mu/D$ versus $\Delta$ for system size $L = 2048$ (values for other
sizes are identical). Lines are a guide for the eye; error bars are smaller than
symbols.}
\end{center}
\end{figure}

In a stochastic system in which the particles are subject a weak force or bias
$f$, we expect the mean displacement per unit time to follow $\langle \Delta x
\rangle/t = \mu f$, where we have assumed the force acts along the $x$
direction, and $\mu$ denotes the mobility. We implement the bias by altering the
probabilities $p(i',j')$ to hop from site $(i,j)$ to a neighbor $(i',j')$ as
follows: $p(i \!-\! 1,j) = 1/4 - f$; $p(i \!+\! 1,j) = 1/4 + f$; $p(i,j \!-\! 1)
= p(i,j \!+\! 1) = 1/4$, with $f \ll 1$. Figure \ref{fig:meanv} (a) shows
$\langle \Delta x (t) \rangle /t \equiv \overline{v}$ as a function of $f$ for
several values of $p$, and system sizes $L = 1024$ and $2048$, in the stationary
regime. Note that $\overline{v}$ is independent of system size, and is
proportional to $f$, as expected.

\begin{figure}[htb]
\begin{center}
\vspace{0.5cm}
\includegraphics[scale=0.29]{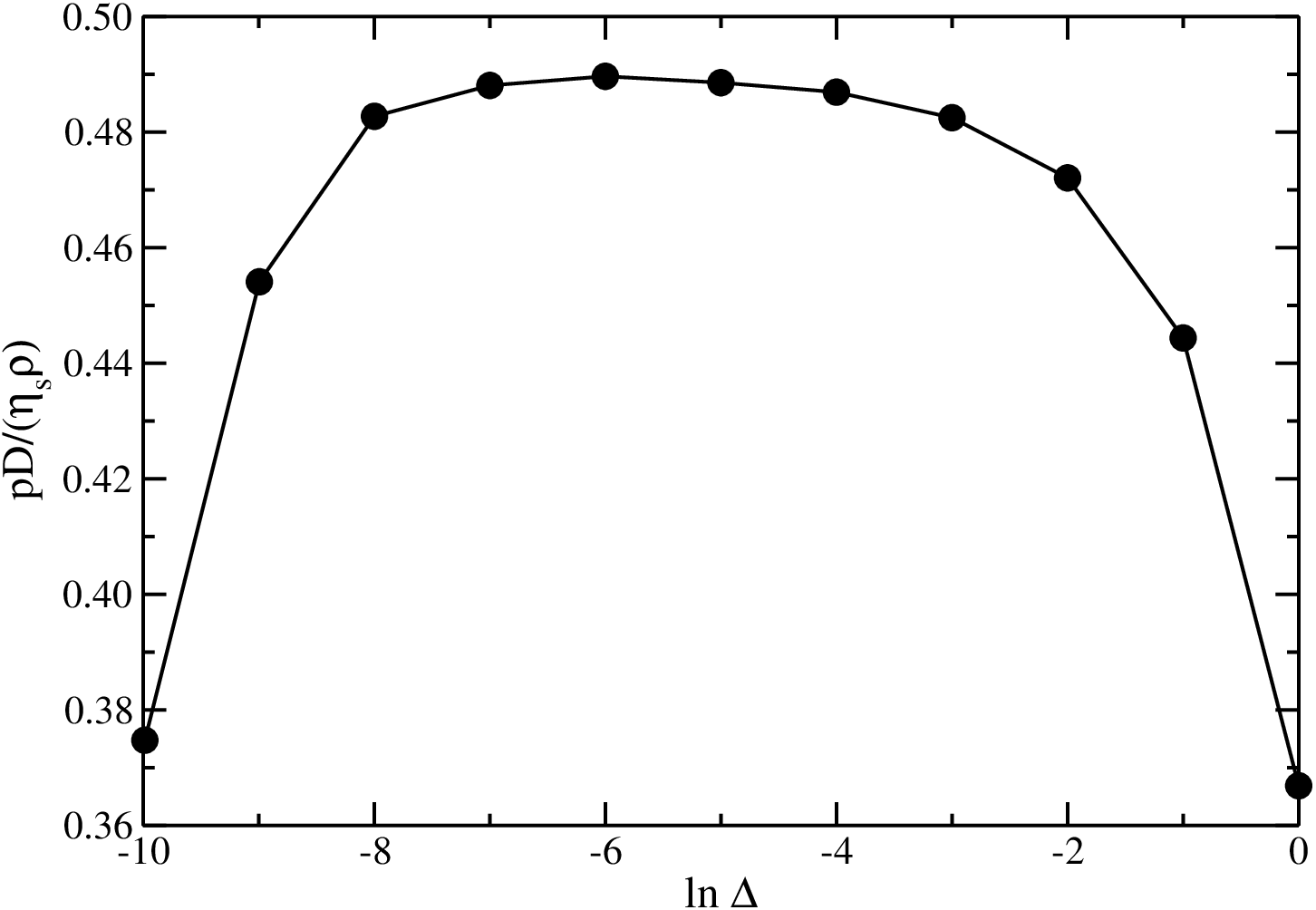}
\caption{\label{fig:Drho} Stationary ratio $pD/(\eta_s \rho)$ versus $\Delta$
for system size L=2048.}
\end{center}
\end{figure}

If particle hopping events were mutually independent, we would expect a
simple relation (analogous to the Stokes-Einstein relation of Brownian motion)
to hold between $\mu$ and $D$. To see this, note that at each particle
displacement, $\langle (\Delta x)^2 + (\Delta y)^2 \rangle = 1$, and that the
fraction of particles residing at active sites is $2 \rho/p$. Thus, assuming
independence, the particle msd (averaged over all particles) during a time
interval $t$ is $2\eta_s \rho t/p$, yielding $D = \rho \eta_s/2p$.  Under a bias
$f$, the mean displacement along the $x$ direction per particle displacement is
$\langle \Delta x \rangle = 2f$, so that $v = 4 \eta_s \rho f/p$, which implies
$\mu/D = 8$.  In fact, the ratio $\mu/D$ remains close to this value over most
of the range of particle densities of interest, as shown in the inset of
Fig.~\ref{fig:meanv}. Finally, the simplistic prediction $pD/(\rho
\eta_s) = 1/2$ is tested in Fig.~\ref{fig:Drho}. We see that the ratio remains
close to, but smaller than, the predicted value, over most of the range of
interest, with more significant deviations neat the critical point and for $p$
approaching 2. The discrepancies between the simplistic predictions for
$\mu/D$ and $pD/(\rho \eta_s)$ and the values observed in simulations presumably
reflect correlations between occupancies of nearby sites.
\vspace{0.5cm}

\section{Conclusions}

In summary, we study diffusion in a restricted-height conserved stochastic sandpile
in two dimensions, determining the critical particle density $p_c$ to high
precision. Our results show that, as expected, the diffusion constant scales in
the same manner as the order parameter.  An analogous result was shown
previously for several one-dimensional conserved stochastic sandpiles
\cite{epjb2009}. At short times we observe subdiffusive behavior, i.e., a msd
growing $\sim t^\gamma$ with $\gamma < 1$. This reflects the steady initial
decay of activity (which follows a power law at criticality, $\rho \sim
t^{-\delta}$) and does not imply anomalous diffusion, since the msd in fact
grows linearly with the number of topplings, so that $\gamma = 1 - \delta$ at
criticality.

\begin{table}[h]
\begin{center}
\begin{tabular}{|c|c|c|} \hline
                    & Langevin & present work  \\ \hline\hline
 $\beta/\nu_\perp$  & 0.85(8)  & 0.77(1)       \\
 $\nu_\perp$        & 0.84(2)  & 0.82(1)      \\
 $z$                & 1.51(3)  & 1.512(8)      \\
 $\beta$            & 0.66(2)  & 0.63(1)      \\
 $\delta$           & 0.50(5)  & 0.43(1)       \\
\hline
\end{tabular}
\end{center}
\caption{Summary of results for critical exponents characterizing the
two-dimensional conserved stochastic sandpile.  Results in the column labeled
``Langevin" are from Ref.~\cite{ramasco}.}
\label{tab1}
\end{table}

Our estimates for critical exponents, summarized in Table I, are consistent with
studies based on the Langevin equation \cite{ramasco}. The somewhat higher
precision of the present study derives in good part from the precision of our
estimate for the critical density, $p_c= 0.7112687(2)$. The associated critical
moment ratio is $m=1.2354(2)$. The activity density and diffusion constant
exhibit a finite-size scaling collapse of data over a restricted range
of particle densities, although somewhat larger than in the one-dimensional
CSS. Overall, our results support a fairly simple scaling picture for the
two-dimensional CSS, including the stationary correlation function for the
activity.  This is in contrast to the one-dimensional case, which is marked by
anomalous scaling behavior, and for which the issue of asymptotic DP-like scaling
remains open.

\section*{Acknowledgements}

This work is supported by CNPq, Brazil.

\end{document}